\documentclass[preprint, 3p]{elsarticle}
\journal{JSV}
\usepackage{lineno,hyperref}
\usepackage{graphicx}           % standard LaTeX graphics tool
\usepackage{amsmath}
\usepackage{amssymb}
\usepackage{xspace}
\usepackage{color}
\newcommand{\nc}{\newcommand}
\nc{\rnc}{\renewcommand}
\nc{\bs}{\boldsymbol}
\rnc{\matrix}[2]{\left[\!\!\begin{array}{#1}
	#2\end{array}\!\!\right]}
\rnc{\vector}[1]{\matrix{c}{#1}}
\nc{\mm}[1]{\boldsymbol{#1}}
\nc{\mms}[1]{\boldsymbol{#1}}
\nc{\real}[1]{\Re\left\{ #1 \right\}}
\nc{\imag}[1]{\Im\left\{ #1 \right\}}
\nc{\dd}{\mathrm{d}}
\nc{\ii}{\mathrm{i}}
\nc{\ee}{\mathrm{e}}
\nc{\inv}{^{-1}} %not used
\nc{\herm}{^{\mathrm H}}
\nc{\tra}{^{\mathrm T}}
\nc{\conj}[1]{ \overline{#1} }
\nc{\MM}{\mm M}
\nc{\BB}{\mm B}
\nc{\LL}{\mm L}
\nc{\ZZ}{\mm Z}
\nc{\eye}{\mm I}
\nc{\qq}{\mm q}
\nc{\dq}{\dot{\qq}}
\nc{\ddq}{\ddot{\qq}}
\nc{\fex}{\mm f}
\nc{\fexx}{\fex_{\mathrm{ex}}}
\nc{\fexxh}{\hat{\fex}_{\mathrm{ex}}}
\nc{\fcs}{c}
\nc{\fc}{\mm \fcs}
\nc{\fh}{\hat \fcs}
\nc{\fhnull}{\fh_0}
\nc{\xx}{\mm x}
\nc{\xr}{\tilde{\xx}}
\nc{\vv}{\mm\varphi}
\nc{\eex}{\mm e_{\fcs}}
\nc{\eextra}{\mm e\tra_{\fcs}}
\nc{\zA}{Z_{A}}
\nc{\zAst}{\zA^*}
\nc{\za}{Z_{\mathrm a}}
\nc{\ndof}{M}
\nc{\ndim}{N}
\nc{\ncon}{C}
\nc{\phiex}{\varphi_{\mathrm{ex}}}
\nc{\qtip}{q_{\mathrm{tip}}}
\nc{\qex}{q_{\mathrm{ex}}}
\nc{\qexh}{\hat q_{\mathrm{ex}}}
\nc{\mmod}{m_{\mathrm{mod}}}
\nc{\COMMENT}[1]{\textcolor{red}{#1}}
\nc{\ie}{i.\,e.\xspace}
\nc{\eg}{e.\,g.\xspace}
\nc{\cf}{cf.\,}
\nc{\myquote}[1]{`#1'}
\nc{\etal}{et al.\xspace}
\nc{\fabstand}{\,}
\nc{\fp}{\fabstand.}
\nc{\fk}{\fabstand,}
\nc{\tab}[5][tbh]{\begin{table}[#1]\centering\caption{#4\label{tab:#5}}\begin{tabular}{#2}\hline #3 \\ \hline\end{tabular}\end{table}}
\nc{\fig}[4][tbh]{
\begin{figure}[#1]
\centering
\includegraphics[width=#4\textwidth]{figures/#2}
\caption{#3\label{fig:#2}}
\end{figure}}
\nc{\e}[2]{\begin{equation} #1 \label {eq:#2} \end{equation}}
\nc{\est}[1]{\begin{equation*} #1 \end{equation*}}
\nc{\ea}[1]{
\begin{eqnarray}
#1\end{eqnarray}}
\nc{\east}[1]{
\begin{eqnarray*}
#1 \end{eqnarray*}}
\nc{\fref}[1]{{Fig.~\ref{fig:#1}}}
\nc{\frefo}[1]{{\ref{fig:#1}}}
\nc{\frefs}[1]{{Figs.~\ref{fig:#1}}}
\nc{\tref}[1]{{Tab.~\ref{tab:#1}}}
\nc{\trefo}[1]{{\ref{tab:#1}}}
\nc{\trefs}[1]{{Tab.~\ref{tab:#1}}}
\nc{\eref}[1]{{Eq.~(\ref{eq:#1})}}
\nc{\erefo}[1]{(\ref{eq:#1})}
\nc{\erefs}[1]{{Eqs.~(\ref{eq:#1})}}
\nc{\sref}[1]{{Section~\ref{sec:#1}}}
\nc{\srefo}[1]{\ref{sec:#1}}
\nc{\srefs}[1]{{Sections~\ref{sec:#1}}}
\nc{\aref}[1]{{{\ref{asec:#1}}}}
\nc{\arefo}[1]{{\ref{asec:#1}}}
\nc{\arefs}[1]{{{Appendices~\ref{asec:#1}}}}

\makeindex             % used for the subject index

\begin{document}

\begin{frontmatter}
\title{Extension of the single-nonlinear-mode theory by linear attachments and application to exciter-structure interaction}
\author{
Malte Krack$^1$
}
\address{$^1$ University of Stuttgart, GERMANY}
%\address{email addresses}

\begin{abstract}
% CONTEXT
Under certain conditions, the dynamics of a nonlinear mechanical system can be represented by a single nonlinear modal oscillator.
This holds, in particular, under external excitation near primary resonance or under self-excitation by negative damping of the respective mode.
The properties of the modal oscillator can be determined by computational or experimental nonlinear modal analysis.
The simplification to a single-nonlinear-mode model facilitates qualitative and global analysis, and substantially reduces the computational effort required for probabilistic methods and design optimization.
% ORIGINAL DEVELOPMENT
Important limitations of this theory are that only purely mechanical systems can be analyzed and that the respective nonlinear mode has to be recomputed when the system's structural properties are varied.
With the theoretical extension proposed in this work, it becomes feasible to attach linear subsystems to the primary mechanical system, and to approximate the dynamics of this coupled system using only the nonlinear mode of the primary mechanical system.
The attachments must be described by linear ordinary or differential-algebraic equations with time-invariant coefficient matrices.
The attachments do not need to be of purely mechanical nature, but may contain, for instance, electric, magnetic, acoustic, thermal or aerodynamic models.
This considerably extends the range of utility of nonlinear modes to applications as diverse as model updating or vibration energy harvesting.
%These linear attachments are generalized linear time-invariant dynamical systems, mechanically connected to the nonlinear mechanical host system.
%Here, \myquote{generalized} means that the attachments do not need to be of purely mechanical nature, the only formal requirement is that they have to be described by a linear ordinary or differential-algebraic equation system.
As long as the attachments do not significantly deteriorate the host system's modal deflection shape, it is shown that their effect can be reduced to a complex-valued modal impedance and an imposed modal forcing term.
% APPLICATION
In the present work, the proposed approach is computationally assessed for the analysis of exciter-structure interaction.
More specifically, the force drop typically encountered in frequency response testing is revisited.
A cantilevered beam with cubic spring and an attached electro-dynamical shaker serves as benchmark.
The proposed approach shows excellent accuracy.
Mainly the already known limitations of single-nonlinear-mode theory reappear.
In particular, higher harmonics should not be too pronounced.
In the transient case, the time scales of vibration and amplitude-phase modulation should be well separated, and the attachment dynamics should be in quasi-steady state.
\end{abstract}


\end{frontmatter}

\section*{Nomenclature}
\begin{eqnarray*}
% SCALARS
t && \text{time}\\
\Omega && \text{angular oscillation frequency}\\
\zA && \text{complex modal impedance}\\
a && \text{modal amplitude}\\
\theta && \text{modal phase lag}\\
\omega && \text{modal frequency}\\
D && \text{modal damping ratio}\\
% VECTORS
\vv_n && k\text{-th complex Fourier coefficient of mode shape}\\
\vv=\vv_1 && \text{fundamental complex Fourier coefficient of mode shape}\\
\qq && \text{vector of generalized coordinates (structure)}\\
\xx && \text{vector of state variables (attachment)}\\
\xr && \text{unique set of state variables (attachment)}\\
\mm g && \text{vector of generalized internal forces (structure)}\\
\fex^{(s)} && \text{vector of input with known time dependence acting on component}\,s\\
\fc^{(s)} && \text{vector of coupling forces acting on component}\,s\\
\fcs,\eex && \text{scalar coupling force and its unit direction vector}\\
\fexxh && \text{imposed force acting on nonlinear modal oscillator}\\
% MATRICES
\eye && \text{identity matrix}\\
\MM && \text{mass matrix (structure)}\\
\mm A_k && k\text{-th order time derivative coefficient matrix (attachment)}\\
\BB && \text{compatibility matrix}\\
\BB^{(s)} && \text{restriction of}\,\BB\,\text{to component}\,s\\
\LL && \text{localization matrix}\\
\LL_m && \text{submatrix of}\,\LL\\
\ZZ && \text{impedance matrix of free attachment}\\
\ZZ_{mn} && \text{impedance submatrices of assembled attachment}\\
% NUMBERS
\ndof && \text{number of generalized coordinates (structure)}\\
\ndim && \text{number of state variables (attachment)}\\
\ncon && \text{number of interface constraints}\\
H && \text{harmonic truncation order}\\
% OPERATORS
\dot{\square} && \text{derivative with respect to}\, t\\
\hat{\square} && \text{fundamental complex Fourier coefficient}\\
\conj{\square} && \text{complex conjugate}\\
\square\tra && \text{(real) transpose}\\
\square\herm && \text{complex-conjugate (Hermitian) transpose}\\
\square^+ && \text{pseudoinverse} %\\
\end{eqnarray*}

%%========================================================================================================================
%% Introduction
%%========================================================================================================================
\section{Introduction}\label{sec:intro}
% Brief recap of nonlinear modes
The concept of normal modes is quintessential to the vibration analysis of linear mechanical systems.
Consequently, scientists have explored opportunities to extend this powerful concept to the nonlinear case.
The earliest attempts date back to the 1960s \cite{rose1960}; for a thorough review article, the reader is referred to \cite{kers2009}.
% Definition
A nonlinear mode is commonly defined as a family of periodic oscillations of a conservative mechanical system (see \eg \cite{vaka2008,kers2009}).
It originates from a corresponding normal mode of the linearized system at the equilibrium point, and continuously extends this with respect to an amplitude measure (\eg mechanical energy).
% Modal analysis
Modal analysis denotes the process of determining modal properties, \ie, modal frequency, damping ratio and deflection shape (including higher harmonics).
These are amplitude-dependent in the nonlinear case.
For numerical nonlinear modal analysis, standard computational techniques can be used, including numerical continuation and bifurcation analysis.
Techniques for experimental nonlinear modal analysis are also available \cite{gibe2003,Peeters.2011,Zapico-Valle2013,Dion.2013b,Ehrhardt.2016,Scheel.2018,Karaagacl.2021,Renson.2016,Renson.2017b}, although it should be remarked that this is still a rather active research area.
\\
% USE OF NONLINEAR MODES; single-nonlinear-mode theory
The amplitude-dependent modal properties provide insight into the qualitative vibration behavior (\eg softening/hardening, importance of higher harmonics, energy localization).
Under some conditions, the vibration energy is confined to a single nonlinear mode (\emph{single-nonlinear-mode theory}) \cite{szem1979}.
In particular, this is the case if the mechanical system is externally driven near a well-separated primary resonance, which can be formally argued using perturbation theory \cite{Cenedese.2019}.
This may also be the case under self-excitation \cite{krac2013a,Jahn.2020}, or combinations of self- and forced excitation \cite{Heinze.2019}.
When the vibration is dominated by a single nonlinear mode, the dynamics take place on a two-dimensional invariant manifold in state space \cite{shaw1991,shaw1993,Haller.2016}, and the mechanical system behaves like a single-degree-of-freedom oscillator.
The governing equation of this nonlinear modal oscillator (or nonlinear-mode model) can formally be derived by nonlinear projection onto the invariant manifold \cite{krac2014a}.
Using Complexification-Averaging, slow modulations of modal amplitude and phase lag can also be considered, \eg, to approximate transient sweeps through resonance or the free decay from resonance \cite{krac2014a}.
%, and the resonant frequency response closely follows the frequency-amplitude relation of the nonlinear mode.
The reduction to a single oscillator facilitates exhaustive parameter studies, probabilistic analyses and design optimization \cite{krac2014b,Sun.2020}.
This also facilitates to use qualitative methods such as averaging and multiple scales analysis, to apply global analysis techniques, and to obtain closed-form expressions for the vibration response.
It is thus possible to gain qualitative understanding of complicated nonlinear phenomena, including the occurrence and disappearance of isolated frequency response branches \cite{Ponsioen.2019}.
In conjunction with either numerical computation or experimental identification of the modal properties, this forms a powerful quantitative-qualitative approach.
The above described model order reduction technique based on nonlinear modes is immediately applicable on system level.
A dynamic substructuring technique based on nonlinear modes was proposed by Thouverez et al. \cite{Joannin.2017,Joannin.2018}.
Starting from the Craig-Bampton method, their idea is to replace the set of linear fixed-interface normal modes (on component level) by a single nonlinear fixed-interface mode.
\\
% SCOPE AND OUTLINE
The purpose of the present work is to extend the single-nonlinear-mode theory by \emph{linear attachments}.
A linear attachment is defined as a generalized (not necessarily purely mechanical) linear time-invariant dynamical system, which is mechanically connected to the nonlinear mechanical host system, as described in \sref{substruct}.
The extension of the nonlinear-mode model is developed in \sref{extend}.
The approach is then applied to exciter-structure interaction in \sref{app}.
The numerical results in \sref{num} show the opportunities and limitations of the proposed approach.
Conclusions are drawn in \sref{conc}.

\section{Mathematical description of structure and attachment, and their coupling}\label{sec:substruct}
% SETTING
The setting illustrated in \fref{schematic} is considered, consisting of two components, a \emph{structure} and an \emph{attachment}, which are coupled to each other.
% STRUCTURE
The structure is here defined as a time-invariant nonlinear mechanical system.
It is assumed that the structure is described in a spatially discrete form, with the system of second-order ordinary differential equations of motion \eref{component1},
\ea{
\MM \ddq + \mm g\left(\qq,\dq\right) &=& \fex^{(1)}(t) + \fc^{(1)} \fk \label{eq:component1}\\
\sum_k \mm A_k \frac{\dd^k}{\dd t^k} \xx &=& \fex^{(2)}(t) + \fc^{(2)} \fp \label{eq:component2}
}
Herein, $\MM=\MM\tra>\mm 0$ denotes the structure's time-constant, symmetric and positive definite mass matrix ($\MM\in\mathbb R^{\ndof\times\ndof}$), $\qq\in\mathbb R^{\ndof\times 1}$ is the vector of $\ndof$ generalized coordinates, overdot denotes derivative with respect to time $t$, and $\mm g\in\mathbb R^{\ndof\times 1}$ is the vector function of internal forces.
The internal forces may contain linear and nonlinear restoring and damping terms.
The nonlinear terms may represent local or global nonlinearities.
\fig[t!]{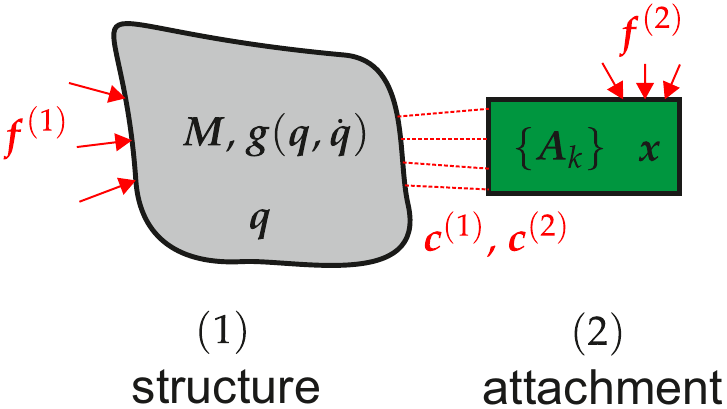}{Schematic illustration of structure, attachment, and their coupling}{0.4}
\\
% ATTACHMENT
\eref{component2} describes the attachment, which is a time-invariant linear generalized dynamical system, with the time-constant coefficient matrices $\mm A_k\in\mathbb R^{\ndim\times\ndim}$ and the vector of $\ndim$ state variables $\xx\in\mathbb R^{\ndim\times 1}$.
Here, generalized means that the attachment does not have to be a purely mechanical model, but it may additionally consist of, \eg, electrical, magnetic, acoustic, thermal or aerodynamic model parts.
\\
% EXTERNAL INPUT + COUPLING
$\fex^{(1)},\fc^{(1)}\in\mathbb R^{\ndof\times 1}$ are external forces acting on the structure.
$\fex^{(1)}$ are applied forces of known time dependence, while the forces $\fc^{(1)}$ are due to the interactions between the structure and the attachment (\fref{schematic}).
%The attachment is coupled to the structure and, thus, it has a mechanical interface.
Since the structure is a strictly mechanical model, the interface between structure and attachment is necessarily mechanical as well.
This means that the interface is described by (mechanical) displacements and the interactions between the structure and attachment are described by (mechanical) forces.
This implies that $\fc^{(2)} \in\mathbb R^{\ndim\times 1}$ also represents forces.
In contrast, $\fex^{(2)}(t)\in\mathbb R^{\ndim\times 1}$ is a generalized input vector with known time dependence.
It may represent imposed forces, currents, magnetic fluxes, and so on.
The coupling forces, $\fc^{(1)}$, $\fc^{(2)}$ can generally not be given as explicit functions of $\qq$ and $\xx$; they are defined in the next subsection.
\\
% COMPARISON OF MATHEMATICAL DESCRIPTION
It is useful to stress the differences of the mathematical models of the structure and the attachment:
While \eref{component1} is nonlinear, \eref{component2} is strictly linear.
While \eref{component1} is an ordinary differential equation system, \eref{component2} may be either an ordinary or differential-algebraic equation system (\ie, the coefficient matrices $\mm A_k$ are not required to be regular).
While \eref{component1} involves time-derivatives of up to second order, \eref{component2} is of arbitrary, yet finite order.

\subsection*{Coupling of structure and attachment}
% MULTIPLE ATTACHMENTS
There can be multiple attachments; all attachments are summarized here for convenience.
In practice, splitting into separate attachments with individual interfaces may be useful to exploit the sparsity in the subsequent matrix algebra.
\\
% SUBSTRUCTURING
The coupling between structure and attachment is formulated in the common \emph{dynamic substructuring} framework \cite{klerk2008}, and readers not familiar with the notation are encouraged to visit this reference for examples that illustrate the compatibility and localization matrices and their relation.
For $\fc^{(1)}=\mm 0$, $\fc^{(2)}=\mm 0$, \eref{component1} and \eref{component2} govern the dynamics of the individual structure and the individual attachment, respectively (without coupling between them).
Since structure and attachment are part of the same assembly, two conditions can be imposed on the interface, compatibility of displacements and force equilibrium.
The compatibility condition is expressed as
\ea{
\underbrace{\matrix{cc}{\BB^{(1)} & \BB^{(2)}}}_{\BB}  \vector{\qq \\ \xx} = \mm 0 \fk \label{eq:compatibility}
}
%
% COMPATIBILITY
where $\BB^{(1)}\in\mathbb R^{\ncon\times\ndof}$, $\BB^{(2)}\in\mathbb R^{\ncon\times\ndim}$ are the so-called compatibility matrices, and $\ncon$ is the number of interface constraints.
$\BB^{(1)}\qq$ and $\BB^{(2)}\xx$ extract the corresponding displacements at either side of the interface, with opposite sign, so that the individual equations of \eref{compatibility} are of the form $u_\Gamma^{(1)}-u_\Gamma^{(2)}=0$.
Typically $\BB^{(s)}$ are signed Boolean and mainly take care of the appropriate sorting.
They can also map between non-conforming meshes and take care of coordinate transforms.
\\
% PRIMAL VARIABLES
Different opportunities are available to solve for the dynamics of the assembly, including primal and dual assembly.
In this work, a \emph{primal assembly} is pursued.
This relies on choosing a set of primal variables that are uniquely defined for the entire system.
In the following, it is assumed that $\ncon \leq \ndim$; \ie, the number of interface displacements must not exceed the number of attachment states.
It is then possible to retain the entire set of generalized structural coordinates, $\qq$, within the set of primal variables.
Retaining $\qq$ preserves the architecture of the structural model, which is crucial for the approach developed in this work, as will become clear later.
With this, the relation between the component variables $\qq$, $\xx$ and the unique set of primal variables takes the form
\ea{
\vector{\qq \\ \xx} = \underbrace{\matrix{cc}{\eye & \mm 0 \\ \LL_{1} & \LL_{2}}}_{\LL} \vector{\qq \\ \xr} \fp \label{eq:primal}
}
Herein, $\xr\in\mathbb R^{\left(\ndim-\ncon\right)\times 1}$ together with $\qq$ form the unique set of primal variables.
Let us define $\delta \xr$ and $\delta \qq$ as infinitesimal virtual variations, compatible with the interface compatibility constraints, but otherwise arbitrary.
In other words, each primal variable can be varied independently without violating the interface compatibility.
Substituting \eref{primal} into \eref{compatibility}, this leads to $\BB\LL~[\delta \qq\tra\, \delta\xr\tra]\tra=\mm 0$.
Since $\delta \qq$ and $\delta \xr$ are arbitrary, this yields the requirement that $\BB\LL = \mm 0$.
A solution of this equation is
\ea{
\LL_{1} = -\left(\BB^{(2)}\right)^+\,\BB^{(1)} \fk
\qquad
\LL_{2} = \operatorname{null} \BB^{(2)} \fp \label{eq:L1L2}
}
Herein, the dimensions of the matrices are $\LL_1\in\mathbb R^{\ndim\times\ndof}$ and $\LL_2\in\mathbb R^{\ndim\times\left(\ndim-\ncon\right)}$.
$\square^+$ denotes the pseudoinverse, such that $\LL_1$ is the minimum-least-squares solution of $\BB^{(2)}\LL_1=\BB^{(1)}$. %, which is equivalent to the left inverse if $\BB^{(2)}$ has full row rank.
$\LL_2$ is an orthonormal basis of the kernel of $\BB^{(2)}$, formed by the right singular vectors of $\BB^{(2)}$ associated with zero singular values.
In a practical implementation, the matrices $\BB^{(1)}$ and $\BB^{(2)}$ would be formulated, and then $\LL_1$, $\LL_2$ would be determined using \eref{L1L2}.
An example is given in \sref{app}.
\\
% EQUILIBRIUM
As shown above, using the primal variables, the interface compatibility condition \erefo{compatibility} is automatically satisfied.
Let us now consider the force equilibrium at the interface.
The interface forces $\fc^{(1)}$ and $\fc^{(2)}$ can be interpreted as the reaction forces needed to ensure compatible displacements at the interface.
A compatible displacement will always be strictly orthogonal to these reaction forces.
Consequently, the virtual action of the interface forces on the coupled system must vanish, $\delta \qq\tra ~\fc^{(1)} ~+~ \delta \xx\tra~ \fc^{(2)} = 0$.
Substituting \eref{primal} and taking into account, again, that $\delta\qq$ and $\delta\xr$ are arbitrary, this leads to
\ea{
\LL\tra \vector{\fc^{(1)} \\ \fc^{(2)}} = \mm 0 \fp \label{eq:equilibrium}
}
%\eref{compatibilityAndEquilibrium}
This equation expresses the equilibrium of forces across the interface.
$\LL\tra\in\mathbb R^{(\ndof+\ndim-\ncon)\times(\ndof+\ndim)}$ is the so-called localization matrix \cite{klerk2008}.
%$\LL\tra\fc$ extracts and adds corresponding forces, so that the individual equations are of the form $c_\Gamma^{(1)}+c_\Gamma^{(2)} = 0$ at the interface, or simply $c_{\Omega}^{(1)}=0$, $c_{\Omega}^{(2)}=0$ away from the interface.
%Typically $\LL$ is Boolean; otherwise what was stated about $\BB^{(s)}$ holds.
%Obviously, the equilibrated forces can be linearly combined, and hence $\LL$ is not uniquely defined.
%This fact is exploited in the following in order to obtain convenient expressions for the assembly.
\\
% PRIMAL ASSEMBLY
Having defined a unique set of primal variables, the interface forces in \eref{component1} and \eref{component2} can be eliminated.
The general procedure is to substitute \eref{primal} into these equations, and to pre-multiply both sides of the equation system (stacked in a vector) by $\LL\tra$, so that the interface forces are eliminated by exploiting \eref{equilibrium}.
Because of the special form of $\LL$ in \eref{primal}, $\qq$ is retained when substituting the relation in \eref{primal}.
Moreover, all terms in \eref{component1} are only pre-multiplied by the identity matrix.
In contrast, $\xx$ in \eref{component2} must be replaced in accordance with the second hyper-row in \eref{primal}, $\xx = \LL_1\qq + \LL_2\xr$.
Pre-multiplying then the resulting equation by either $\LL\tra_1$ or $\LL\tra_2$ yields
\ea{
\sum_k \LL\tra_{1}\mm A_k \LL_{1} \frac{\dd^k}{\dd t^k} \qq + \LL\tra_{1} \mm A_k \LL_{2} \frac{\dd^k}{\dd t^k} \xr & = & \LL\tra_{1}\fex^{(2)}(t) - \fc^{(1)} \fk \label{eq:assemblyT1} \\
\sum_k \LL\tra_{2}\mm A_k \LL_{1} \frac{\dd^k}{\dd t^k} \qq + \LL\tra_{2} \mm A_k \LL_{2} \frac{\dd^k}{\dd t^k} \xr & = & \LL\tra_{2} \fex^{(2)}(t) \fp \label{eq:assemblyT2}
}
Herein, it is used that $\LL\tra_2\fc^{(2)}=\mm 0$, and $\fc^{(1)}+\LL\tra_{1}\fc^{(2)}= \mm 0$ (\cf interface equilibrium in \eref{equilibrium}).
Finally, to eliminate $\fc^{(1)}$, and thus to complete the primal assembly, one adds both sides of \eref{component1} and \eref{assemblyT1},
\ea{
\MM \ddq + \mm g\left(\qq,\dq\right) + \sum_k \LL\tra_{1}\mm A_k \LL_{1} \frac{\dd^k}{\dd t^k} \qq + \LL\tra_{1} \mm A_k \LL_{2} \frac{\dd^k}{\dd t^k} \xr &=& \fex^{(1)}(t) + \LL\tra_{1}\fex^{(2)}(t) \fk \label{eq:finalassembly1}\\
\sum_k \LL\tra_{2}\mm A_k \LL_{1} \frac{\dd^k}{\dd t^k} \qq + \LL\tra_{2} \mm A_k \LL_{2} \frac{\dd^k}{\dd t^k} \xr & = & \LL\tra_{2} \fex^{(2)}(t) \fp \label{eq:finalassembly2}
}
%
%The $\LL\tra_n\mm A_k\LL_m$ are blocks of the assembled coefficient matrices.
The resulting system of equations \erefo{finalassembly1}-\erefo{finalassembly2} governs the dynamics of the assembly (structure + attachment), and can be numerically solved.
This serves as reference for the following nonlinear-mode-based approximation.

\section{Extension of the single-nonlinear-mode theory by linear attachments}\label{sec:extend}
The idea of this section is to assume that the structural response is dominated by a single nonlinear mode, and thus to reduce \eref{component1} to a single nonlinear modal oscillator subjected to external imposed and interface forces, $\fex^{(1)}$ and $\fc^{(1)}$, respectively.
To determine the interface forces, $\fc^{(1)}$, acting on the modal oscillator, the linear attachment model in \eref{component2} is cast into the frequency domain.
It will be shown that this permits to express the interface forces, projected onto the nonlinear mode, in closed form, as complex impedance and imposed external force of the modal oscillator.
%The linear attachment is described in the frequency domain, which permits to derive a closed-form expression of the coupling forces.
%The attachment is then seen to impose an external force and add a complex impedance to the nonlinear modal oscillator.

\subsection{Nonlinear modal analysis and reduction to a single modal oscillator}
% MODAL PROPERTIES + DEFINITIONS OF NONLINEAR MODES
The first step is to reduce \eref{component1} to a single nonlinear modal oscillator.
The properties of this modal oscillator are the modal frequency $\omega(a)\in \mathbb R$, modal damping ratio $D(a)\in \mathbb R$ and the modal deflection shape (defined later), each of which depends on the modal amplitude $a$.
These modal properties are determined by nonlinear modal analysis of the structure described by \eref{component1}, but without external forces ($\fex^{(1)}=\mm 0 = \fc^{(1)}$).
Various definitions are available for nonlinear modes, and numerous methods are known to determine the amplitude-dependent modal properties.
The Extended Periodic Motion Concept (EPMC) introduced in \cite{Krack.2015a} proposes one way to define a nonlinear mode.
As opposed to most alternatives, this definition rigorously accounts for damping, and it has been shown to be useful to derive techniques for both computational and experimental nonlinear modal analysis.
Hence, the EPMC is used in this work and a brief recap is given in the following paragraph.
It should, however, be noted that some other definitions of nonlinear modes are compatible with approach developed in this work.
\\
% EPMC
The EPMC is limited to nonlinear modes originating from a certain linear mode with nonzero and distinct natural frequency (no repeated eigenvalue).
The definition considers the autonomous form of \eref{component1} without external forces ($\fex^{(1)}$, $\fc^{(1)}$).
A nonlinear mode is defined as the family of periodic solutions of \cite{Krack.2015a}
\ea{
\MM \ddq-2D\omega\MM\dq+\mm g(\qq,\dq) = \mm 0\fk \label{eq:EPMC}
}
that continues the corresponding linear mode to high amplitudes.
The artificial negative damping term $-2D\omega\MM\dq$ compensates the natural dissipation (in average over a vibration cycle).
The EPMC simplifies to the conventional periodic motion concept in the conservative case (where $D=0$), and is consistent with the linear case under modal damping \cite{Jahn.2019b}.
If, at the same time, more than one linear mode shape contributes strongly to the nonlinear mode, and damping is not light (\ie $\left|D\right|$ not $\ll1$), the artificial term will generally distort the modal coupling \cite{Krack.2015a}.
The modal analysis can be carried out numerically, using standard methods for computing periodic oscillations, such as Harmonic Balance or Shooting.
Typically, numerical path continuation is used to compute the branch of periodic solutions, starting from the linear mode at low vibration level, up to the desired vibration level.
The vibration level is commonly measured by the mechanical energy.
Details on the numerical implementation can be found, \eg, in \cite{Krack.2015a,Krack.2019}.
The modal analysis can also be carried out experimentally using feedback-control, \eg in the form of phase-resonant testing \cite{Scheel.2018}, response-controlled step sine testing \cite{Karaagacl.2021} or Control-Based Continuation \cite{Renson.2016b,Renson.2017b}.
For each vibration level, one obtains the modal frequency $\omega$, modal damping ratio $D$ and a periodic oscillation.
For the subsequent formulations, it is useful to express the periodic oscillation as a Fourier series,
\ea{
\qq(t) = \real{\sum_{n=0}^{H} \hat{\qq}_n\ee^{\ii n\omega t}}\fk \label{eq:fourier}
}
with the complex Fourier coefficients $\qq_n\in\mathbb C^{\ndof\times 1}$.
It should be stressed that this does not imply that Harmonic Balance must be used for modal analysis.
Indeed, regardless what analytical, computational or experimental method was used for modal analysis, a simple Fourier analysis is capable of determining the Fourier coefficients $\hat{\qq}$.
In computational and experimental practice, only a finite number of $H$ Fourier coefficients can be determined, as indicated in \eref{fourier}.
For convenience, a modal amplitude $0<a\in \mathbb R$ is introduced and defined by $a = \sqrt{\hat{\qq}\herm_1\MM\hat{\qq}_1}$, where $\square\herm$ denotes Hermitian transpose.
One can then define the harmonics of the modal deflection shape, $\vv_n(a)$, by the relation $\qq_n = a\vv_n(a)$.
Note that due to the definition of the modal amplitude, the fundamental harmonic of the mode shape is mass-normalized, $\vv\herm_1 \MM\vv_1 = 1$.
Note also that $\vv_n$ are generally complex-valued, and thus can represent non-trivial phase lags among the generalized coordinates.
The abbreviation $\vv = \vv_1$ is used in the following.
\\
% REDUCTION
It is next assumed that the response of the structure \emph{with} external forces ($\fex^{(1)}$, $\fc^{(1)}$) is dominated by a single nonlinear mode (\emph{single-nonlinear-mode theory}) of the structure \emph{without} external forces.
As mentioned before, this scenario is typical under near-resonant excitation, self-excitation or combinations thereof \cite{Heinze.2019}, provided that strong nonlinear modal interactions are absent.
In these cases, the structure behaves like a nonlinear single-degree-of-freedom oscillator.
The periodic oscillation of the structure can then be approximated as \cite{krac2014a}
\ea{
\qq(a(t),\theta(t))
= \real{ a\sum_{n=0}^{H} \vv_n(a)\ee^{\ii n\left(\int_0^t \Omega\dd \overline t + \theta\right)} }
%= \real{ \sum_{n=0}^{H} \hat \qq_n\ee^{\ii n\int \Omega\dd t} }
\label{eq:ansatz}
\fp
}
A comparison with \eref{fourier} reveals that the same Fourier coefficients are used in the approximation.
However, the oscillation frequency, $\Omega\in\mathbb R$ is generally allowed to deviate from the natural frequency, and a phase lag $\theta\in\mathbb R$ is considered.
In the autonomous case, the oscillation frequency is set equal to the natural frequency, $\Omega=\omega(a)$.
In the externally-driven case, the oscillation frequency is set equal to the imposed excitation frequency.
In either case, $\Omega$ may vary with time, which is accounted for in \eref{ansatz} by taking the integral.
The new unknowns are now the modal amplitude $a$ and the phase lag $\theta$.
These are assumed to modulate slowly as compared to the oscillation with $\Omega$.
This permits to apply Complexification-Averaging.
A governing equation for $a$ and $\theta$ can be determined by substituting the ansatz (\eref{ansatz}) into \eref{component1}, and requiring orthogonality of the residual with respect to the fundamental harmonic of the nonlinear mode \cite{krac2014a}.
Applied to \eref{component1}, this leads to the following complex nonlinear first-order ordinary differential equation for $a$ and $\theta$,
\ea{
2\ii \Omega\left(\dot a +\ii a\dot\theta\right) + \left(~-\Omega^2 + 2D\left(a\right)\omega\left(a\right)\, \ii\Omega + \omega^2\left(a\right)~\right)a = \left(\vv\left(a\right)\right)\herm\left( \hat{\fex}^{(1)} + \hat{\fc}^{(1)} \right)\ee^{-\ii\theta} \fp \label{eq:NMpre}
}
$\hat{\fex}^{(1)} $ and $ \hat{\fc}^{(1)}$ are the fundamental Fourier coefficients of the imposed and coupling forces, respectively.
The steady limit state is simply obtained by setting $\dot a = 0 = \dot\theta$, which results in an algebraic equation.
As mentioned before, the transient case is interesting to describe, for instance, the free decay from a near-resonant response, or the beating of the amplitude envelope as encountered during a frequency sweep through resonance.
The solution of either the differential or the algebraic form of the equation is discussed later.
\\
% COMMENTS ON FUNDAMENTAL HARMONIC PROJECTION
Limiting the above described nonlinear projection to the fundamental harmonic may lead to inaccuracy if higher harmonics play an important role.
The benefit of this is that the projected nonlinear terms can be expressed in closed form (called \emph{spectral substitution} in \cite{Joannin.2017,Joannin.2018}); \ie, the nonlinear terms in \eref{NMpre} are are readily available from the nonlinear modal analysis.
It should be emphasized that the limitation of the projection to the fundamental harmonic does neither mean that the nonlinear interaction of fundamental and higher harmonics is neglected in the nonlinear modal analysis step, or that one obtains only an approximation of the fundamental harmonic of the response.
\\
% LIMITATIONS
Since the modal properties are determined from the structure-only-configuration, the single-nonlinear-mode approximation may be inaccurate if the attachment significantly deteriorates the modal deflection shape.
Also, the attachment should not generate strong modal interactions, as these are neglected in the single-nonlinear-mode theory.
In particular, such modal interactions can be expected if the assembled linearized system is close to an internal resonance condition,
\eg if there is a closely spaced natural frequency, or a natural frequency being a low-integer multiple of the considered modal frequency.
In the transient case, it is additionally assumed that the amplitude and phase modulation is slow as compared with the oscillation, as required to separate the time scales via averaging \cite{vaka2008b}. %such that the time scales can be separated via averaging.
%\\
%It now becomes clear why retaining $\qq$ as primal variables is crucial: The nonlinear modal analysis
%This subsection is a recap of the theoretical developments in \cite{krac2014a,Krack.2015a} and their application to the structural component described by \eref{component1}.
% $\vv=\vv_{j}^{\mathrm{lin}}$, $\omega=\omega_{j}^{\mathrm{lin}}$, $D = D_{j}^{\mathrm{lin}}$ at low amplitudes $a\to 0$, and continuing to high amplitudes $a$.

\subsection{Derivation of the coupling forces}
% Assumption: quasi-steady attachment dynamics
In the following, a closed-form expression is derived for the coupling forces $\hat{\fc}^{(1)}$, by exploiting the linearity of the attachment model and operating in the frequency domain.
For the \emph{transient case}, it is assumed that the attachment dynamics take place on a faster time scale than the modal amplitude and phase modulation.
Then the \emph{attachment dynamics} can be considered to be in \emph{quasi-steady state}.
For a given attachment model, this can be checked by an eigenvalue analysis and comparing the characteristic frequencies with the dynamics of interest.
If the attachment dynamics cannot be considered to be in quasi-steady state, one has to apply Complexification-Averaging to the attachment model, too, in order to capture the slow modulation of its states.
This leads to a potentially much larger equation system.
In contrast, assuming quasi-steady-state attachment dynamics permits to condense the modulation equations to the single equation \erefo{NMpre} of the nonlinear modal oscillator, which is an important simplification.
It is important to emphasize that this assumption is only relevant for the transient case.
In the steady limit state, of course, the equations of both the attachment and the nonlinear modal oscillator do not contain any modulation terms (terms like $\dot a$, $\dot\theta$).
\\
% From time to frequency domain
Assuming steady-state conditions, \erefs{assemblyT1}-\erefo{assemblyT2} can be cast into the frequency domain,
\ea{
\ZZ_{11} \hat\qq + \ZZ_{12} \hat \xr &=& \LL\tra_{1} \hat{\fex}^{(2)} - \hat{\fc}^{(1)}\fk \label{eq:assemblyF1} \\
\ZZ_{21} \hat\qq + \ZZ_{22} \hat \xr &=& \LL\tra_{2} \hat{\fex}^{(2)}\fp \label{eq:assemblyF2}
}
Herein, $\hat{\square}$ denotes the fundamental Fourier coefficient of $\square$, and $\ZZ_{mn} = \LL\tra_m \ZZ\LL_n$, where $\ZZ = \sum_k\left(\ii\Omega\right)^k\mm A_k\in\mathbb C^{\ndim\times\ndim}$ is the impedance matrix of the free attachment.
$\ZZ_{mn}$ are the blocks of the assembled attachment impedance.
The dimensions are
$\ZZ_{11}\in\mathbb C^{\ndof\times\ndof}$,
$\ZZ_{12}\in\mathbb C^{\ndof\times\left(\ndim-\ncon\right)}$,
$\ZZ_{21}\in\mathbb C^{\left(\ndim-\ncon\right)\times\ndof}$,
$\ZZ_{22}\in\mathbb C^{\left(\ndim-\ncon\right)\times\left(\ndim-\ncon\right)}$.
\\
The linear algebraic equation system \erefo{assemblyF1}-\erefo{assemblyF2} can be solved for the coupling forces,
\ea{
\hat{\fc}^{(1)} = - \underbrace{\left(\ZZ_{11}-\ZZ_{12}\ZZ\inv_{22}\ZZ_{21}\right)}_{\ZZ_q}\hat{\qq} + \left(\LL\tra_1-\ZZ_{12}\ZZ\inv_{22}\LL\tra_2\right)\hat{\fex}^{(2)}\fp \label{eq:fc1}
}
Recall that $\ncon \leq \ndim$ was assumed.
If indeed  $\ncon = \ndim$, \eref{assemblyF2} vanishes, all terms including the index $2$ need to be removed in \eref{fc1} and the following.
If $\ncon < \ndim$, it needs to be assumed that the square complex matrix $\ZZ_{22}$ is regular in the considered frequency range.
This corresponds to requiring that the attachment has no (undamped) resonance under fixed-interface conditions $\qq=\mm 0$.
This is not really an additional limitation in practice since such an internal resonance condition has already been excluded by assuming the absence of strong modal interactions.
\\
It should be noted that \erefs{assemblyF1}-\erefo{fc1} hold for arbitrary $\Omega$, not only with respect to the fundamental harmonic.
This can be useful to recover the higher harmonics of the state variables $\xr$ and the coupling forces generated by the nonlinear modal oscillator in the post processing.
\\
% Modal projection
Finally, substituting \eref{fc1} into \eref{NMpre} and using $\hat\qq = a\vv(a)\ee^{\ii\theta}$ yields
\ea{
2\ii \Omega\left(\dot a +\ii a\dot\theta\right) + \left(~-\Omega^2 + 2D\left(a\right)\omega\left(a\right)\, \ii\Omega + \omega^2\left(a\right) + \zA\left(a,\Omega\right) ~\right)a = \left(\vv\left(a\right)\right)\herm \fexxh\left(\Omega\right)\ee^{-\ii\theta} \fk \label{eq:NM}
}
with $\zA(a,\Omega) =  \left(\vv\left(a\right)\right)\herm\ZZ_q(\Omega)\vv\left(a\right)\in\mathbb C$ and $\fexxh(\Omega) = \hat{\fex}^{(1)} + \left(\LL\tra_1-\ZZ_{12}\ZZ\inv_{22}\LL\tra_2\right)\hat{\fex}^{(2)}\in\mathbb C^{\ndof\times 1}$.
Apparently, the attachment introduces an imposed force and a complex modal impedance $\zA(a,\Omega)$ to the nonlinear modal oscillator.
In many cases, the mode shape $\vv(a)$ depends only weakly on the modal amplitude $a$.
Then also $\zA(a,\Omega)$ and the projected imposed forces depend only weakly on $a$.
\\
% ANALYTICAL/NUMERICAL SOLUTION
\eref{NM} can be easily split into real and imaginary parts to obtain a set of two real first-order ordinary differential equations.
These equations can be brought into explicit form with respect to $\dot a$ and $\dot\theta$ \cite{krac2014a}.
As assumed, the modulation takes place on a slower time scale than the oscillation.
Moreover, the nonlinear terms vary rather smoothly with the amplitude $a$.
Consequently, the numerical integration is typically possible with standard methods, and using relatively large time steps.
For this, the nonlinear terms $\omega(a)$, $D(a)$ and $\vv(a)$ must be available as continuous functions.
Piecewise cubic Hermite interpolation between the discrete points obtained by numerical or experimental modal analysis has been shown to give very good results.
For the steady state, it is useful to note that the phase lag $\theta$ can be eliminated from \eref{NM} by taking the magnitude on both sides.
This simplifies the problem to finding the roots of a scalar function in a scalar variable ($a$).
Note that $\fexxh$ and $\zA$ are rational functions in $\Omega$.
It can thus be useful to solve for $\Omega$ (instead of solving for $a$) to determine the frequency-response.
For the special case where $\zA=0$ and $\fexxh$ is frequency-independent, the closed-form solution is given in \cite{Schwarz.2019}.
\\
% USE OF EXPERIMENTAL DATA
Finally, it is useful to note that a model of the structure, in the form of \eref{component1}, is not necessarily needed to set up the single-nonlinear-mode model in \eref{NM}.
Indeed it is an interesting prospect to determine the modal properties experimentally, using nonlinear modal testing, and to use either an experimentally obtained frequency-domain representation, or a theoretical model of the attachment(s).

\section{Application to exciter-structure interaction}\label{sec:app}
The approach developed in the previous section is now applied to analyze the interaction between a nonlinear structure and an attached vibration exciter (shaker).
The structure under test is a cantilevered beam with cubic spring at its free end, see \fref{benchmark_model}a.
The setup and parameter setting is similar to the ECL benchmark \cite{Thouverez.2003}.
This defines the structural model in terms of $\qq$, $\MM$, $\mm g$, as specified later.
The electro-mechanical model of the exciter is illustrated in \fref{benchmark_model}b.
The \emph{free} attachment is described by the linear ordinary equation system \cite{McConnell.1995}
\ea{
L\dot I + R I + G\dot q_{\mathrm a} &=& u(t) \fk \\
m_{\mathrm a}\ddot q_{\mathrm{a}} + d_{\mathrm a}\dot q_{\mathrm a} + \left(k_{\mathrm a}+k_{\mathrm s}\right)q_{\mathrm a} - k_{\mathrm s} q_{\mathrm s} &=& GI\fk \\
k_{\mathrm s} q_{\mathrm s} - k_{\mathrm s}q_{\mathrm a} &=& 0\fp
}
Herein, $L$ and $R$ are the electrical inductance and resistance, $G$ is the force generating constant, $I$ is the electrical current, $q_{\mathrm a}$ is the armature displacement, $q_{\mathrm s}$ is the end point of the stinger, $k_{\mathrm s}$ is the stiffness of the stinger (push rod), and $m_{\mathrm a}$, $d_{\mathrm a}$, $k_{\mathrm a}$ are the armature mass, damping and stiffness.
The link between coil and table is assumed as rigid such that both can be considered as one body (armature), which is a common simplification \cite{McConnell.1995}.
The exciter is operated in voltage mode; \ie, the voltage $u(t)$ is prescribed.
A harmonic input $u(t) = \real{\hat u \ee^{\ii\int_0^t\Omega \dd \overline t}}$ with fixed magnitude but time-varying frequency $\Omega(t)$ is considered.
Note that the above model describes the free attachment without any interface forces.
The interface force $c$ (\fref{benchmark_model}b) arises from the subsequent assembly.
\fig[t!]{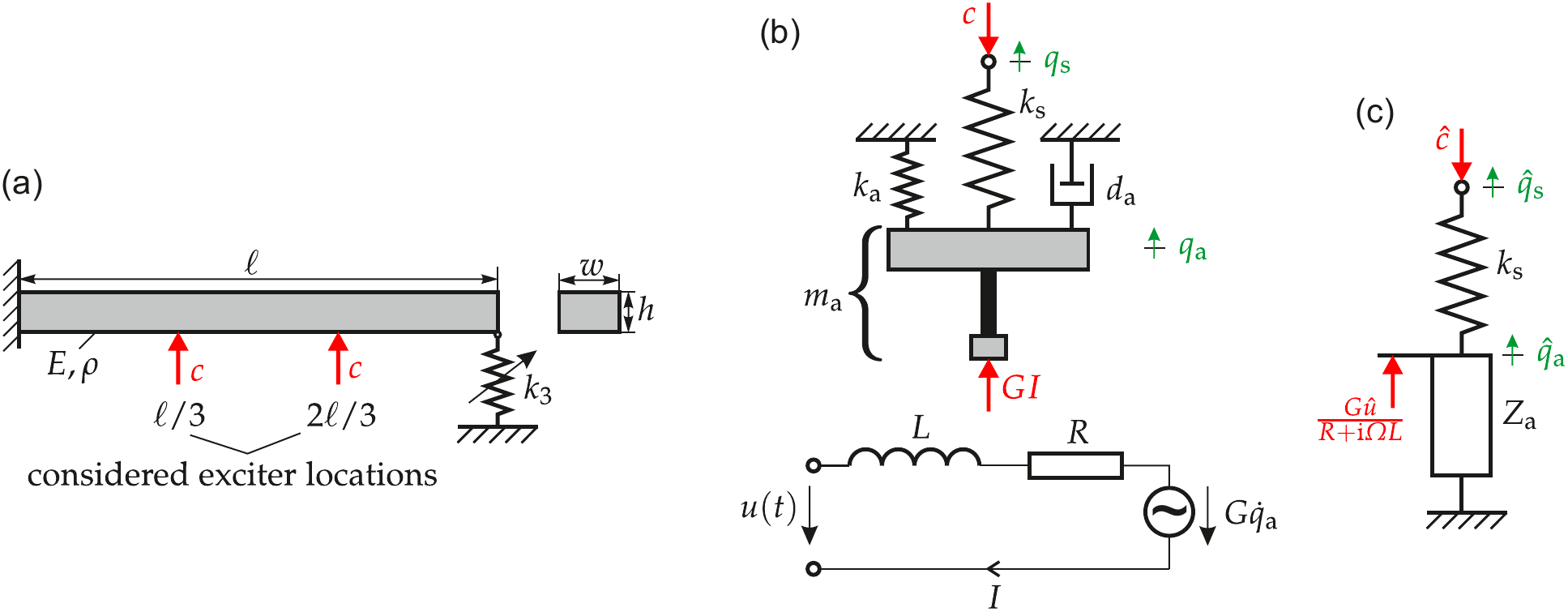}{Model for the analysis of exciter-structure interaction: (a) structure under test, (b) electro-mechanical model of exciter, (c) frequency-domain equivalent model of (b)}{1.0}
\\
The structure is driven only by the exciter ($\hat{\fex}^{(1)}=\mm 0$).
The exciter will later be applied either at $\ell/3$ or at $2\ell/3$ to drive the structure under test with a point load (\fref{benchmark_model}a).
The interface force $\fc^{(1)}$ can thus be expressed as $\fc^{(1)} = \fcs \eex$ where $\fcs\in\mathbb R$ is the scalar applied force and $\eex\in\mathbb R^{\ndof\times 1}$ is its unit direction vector expressed in the structure's generalized coordinates.
In the nomenclature introduced in \sref{substruct}, the attachment states $\xx$, the compatibility matrices and the blocks of the localization matrix read,
\ea{
\xx = \vector{I\\q_{\mathrm a} \\ q_{\mathrm s}}\fk
\qquad
\BB^{(1)} = -\eextra\fk
\quad
\BB^{(2)} = \matrix{ccc}{0 & 0 & 1}\fk
\\
\LL_1 = -\vector{0 \\ 0 \\ 1} \BB^{(1)}\fk
\qquad \LL_2 = \matrix{cc}{1 & 0\\ 0 & 1\\ 0 & 0}\fp
}
Comparing with \eref{compatibility}, one may note that the compatibility condition $\BB^{(1)} \qq + \BB^{(2)} \xx = 0$ means that the end point of the stinger and the force application point on the structure have the same displacement.
$\LL_1$ and $\LL_2$ can be formally calculated using \eref{L1L2}.
One can easily verify that using $\LL_1$ and $\LL_2$ expressed above yields a unique set of primal variables.
\\
The impedance matrix can be identified from the model as
\ea{
\ZZ(\Omega) = \matrix{ccc}{
R+\ii\Omega L & \ii\Omega G & 0\\
-G & -\Omega^2m_{\mathrm a}+\ii\Omega d_{\mathrm a} + k_{\mathrm a} + k_{\mathrm s} & -k_{\mathrm s} \\
0 & -k_{\mathrm s} & k_{\mathrm s}
}\fp
}
Substitution into \eref{fc1} yields
\ea{
\hat{\fc}^{(1)} &=&
\underbrace{\left(~\frac{k_{\mathrm s}}{\za ~ + ~ k_{\mathrm s}} \frac{G\hat u}{R+\ii\Omega L} - \frac{k_{\mathrm s}\za}{\za + k_{\mathrm s}} \left(\eextra \vv(a)a\ee^{\ii\theta}\right)~\right)}_{\hat{\fcs}}\eex \fk\\
&& \underbrace{\frac{k_{\mathrm s}}{\za + k_{\mathrm s}} \frac{G\hat u}{R+\ii\Omega L} \eex}_{{\fexxh}} - \underbrace{\frac{k_{\mathrm s}\za}{\za + k_{\mathrm s}} \eex \eextra}_{\ZZ_q} \underbrace{\vv(a) a\ee^{\ii\theta}}_{\hat{\qq}} \fk
}
with the armature impedance $\za = -\Omega^2m_{\mathrm a}+\ii\Omega d_{\mathrm a} + k_{\mathrm a} + \ii\Omega G^2/\left(R+\ii\Omega L\right)$.
As alternative to the formal derivation with primal assembly, one can interpret the exciter model as the serial connection of armature impedance and stinger stiffness, subjected to a voltage-imposed force (\fref{benchmark_model}c).

\subsection{Simplified expression of the force drop in the linear steady state}
% Context
It is well-established theoretically and experimentally that the applied force $\fcs$ drops in the neighborhood of a structural resonance for an imposed voltage level $\hat u$ \cite{Tomlinson.1979,McConnell.1995,Varoto.2002,Varoto.2002b}.
It is useful for the interpretation of the numerical results to know the conditions leading to a prominent force drop.
Therefore, an expression of the force drop is analytically derived in the following.
The discussion is restricted to the linear case under steady-state conditions.
\\
In the linear case, one has constant modal properties  $\omega = \omega_0$, $D=D_0$ and $\vv=\vv_0$. % D=const, \phi = const
Under steady-state conditions, \eref{NM} then simplifies, after some algebraic manipulations, to
\ea{
\left(-\eta^2 + \ii\eta 2D_0 + 1 + \zAst\right) a = \frac{k_{\mathrm s}}{k_{\mathrm s}+\za}   \frac{1}{1+\ii \eta \frac{\omega_0}{\omega_{\mathrm c}}}  \frac{\conj{\phiex}G\hat u}{\omega_0^2 R} \ee^{-\ii\theta} \label{eq:LMss}
\fp
}
% normalized time $\tau = \omega_0 t$
Herein, $\eta = \Omega/\omega_0$ is the frequency ratio, $\omega_{\mathrm c} = R/L$ is the electrical cutoff frequency, $\phiex = \eextra\vv$, $\conj{\square}$ denotes the complex-conjugate and the normalized attachment impedance $\zAst = \frac{k_{\mathrm s}\za}{\za + k_{\mathrm s}}\left|\phiex\right|^2/\omega_0^2$.
% \\
\eref{LMss} can be solved for $a\ee^{\ii\theta}$.
For the applied force $\fh$, one thus obtains, after some algebraic manipulations:
\ea{
\fh = \frac{k_{\mathrm s}}{\za+k_{\mathrm s}} \frac{1}{1+\ii \eta \frac{\omega_0}{\omega_{\mathrm c}}}  \frac{G\hat u}{R} \left[1 - \frac{k_{\mathrm s}\za}{\za + k_{\mathrm s}} \frac{\left|\phiex\right|^2}{\omega_0^2\left(-\eta^2+\ii\eta 2D_0 + 1 + \zAst\right)} \right]\fp \label{eq:fhlinear}
}
To further simplify this expression, it is assumed that the cutoff frequency is much larger than the modal frequency, $\omega_{\mathrm c}\gg \omega_0$, and the stinger stiffness is much larger than the armature stiffness, $k_{\mathrm s}\gg k_{\mathrm a}$.
As alternative to the latter assumption, one can assume that the exciter armature is directly attached to the structure (without stinger).
%--> armature resonance frequency (structure not moving; new resonance as compared to the system with rigidly attached shaker) >> modal frequency
Under these assumptions, the normalized attachment impedance simplifies to
\ea{
\zAst = \frac{\za\left|\phiex\right|^2}{\omega_0^2} = \mu \left(~-\eta^2+\ii\eta 2 D_{\mathrm a} \frac{\omega_{\mathrm a}}{\omega_0} + \left(\frac{\omega_{\mathrm a}}{\omega_0}\right)^2~\right) \fk
}
with the mass ratio $\mu = m_{\mathrm{a}}/\mmod$, the modal mass (with respect to exciter location) $\mmod = 1/\left|\phiex\right|^2$, and the armature properties $\omega_{\mathrm a} = \sqrt{k_{\mathrm a}/m_{\mathrm a}}$, and
$2D_{\mathrm a}\omega_{\mathrm a} = \left(d_{\mathrm a} + G^2/R\right)/m_{\mathrm a}$.
The term $G^2/R$ corresponds to the electro-mechanical damping.
A larger velocity $\dot{q}_{\mathrm a}$ leads to a larger voltage $G\dot{q}_{\mathrm a}$, which reduces the voltage $RI$ at the resistor, leading to a smaller electrical current $I$ and, thus, a smaller force $GI$ (\cf \fref{benchmark_model}b).
This introduces electro-mechanical damping of viscous type, which may well exceed the mechanical armature damping ($d_{\mathrm{a}}$, assumed viscous here).
\\
With the above stated assumptions, %and abbreviations,
\eref{fhlinear} simplifies to
\ea{
\frac{\fh}{\fhnull} = \frac{-\eta^2 + 2D_0 \ii\eta + 1}{ -\left(1+\mu\right)\eta^2 + 2\left(D_0+\mu \frac{\omega_{\mathrm a}}{\omega_0} D_{\mathrm{a}}\right)\ii\eta + 1 + \mu\frac{\omega_{\mathrm a}^2}{\omega_0^2} } \fk \label{eq:fhfhNull}
}
with $\fhnull = G\hat u/R$.
The right-hand-side of \eref{fhfhNull} is the ratio between the dynamic stiffness of the free structure under test and that of the structure with attached (passive) exciter.
One can verify that $\qexh/\fh$ with $\qexh = \phiex a\ee^{\ii\theta}$, using the solution of \eref{LMss}, yields the expected drive-point frequency response function of the structure under test, $1/\left(-\eta^2 + 2D_0 \ii\eta + 1\right)$.
In contrast,  $\qexh/\fhnull$ is the drive-point frequency response function of the structure with attached (passive) exciter.
Compared to the transfer function of the structure under test, this transfer function is distorted by the added mass, stiffness and electro-mechanical damping of the attached exciter.
Consequently, it is crucial to measure the applied force $\fh$ in a vibration test, as is well-known. % \cite{McConnell}.
\\
The strongest force drop can be expected at the structural resonance, $\eta=1$, where one obtains
\ea{
\left.\frac{\left|\fh\right|}{\left|\fhnull\right|}\right|_{\eta=1} = \frac{2D_0}{\sqrt{ \mu^2\left(\frac{\omega_{\mathrm a}^2}{\omega_0^2} - 1\right)^2 + 4\left(D_0+ \mu\frac{\omega_{\mathrm a}}{\omega_0}D_{\mathrm a}\right)^2 }} \fp \label{eq:forcedrop}
}
For small mass ratios, $0\leq\mu\ll 1$, \eref{forcedrop} yields a value close to unity, such that the force drop is not significant.
In practice, this can be achieved by attaching the exciter closer to the clamping, where $\phiex$ is smaller and thus the modal mass $\mmod$ is larger.
Of course, a larger voltage and force level is then needed to reach the same off-resonant response level.
For moderate mass ratios, the force drop can be considerable, depending on the dynamic properties of the exciter in general, and the electro-mechanical damping in particular.
It should be emphasized that the force drop described here is \emph{not a transient effect}.
If a constant force level is desired, feedback control can be used.
%
%nominal parameters in benchmark: $D_0=0.4\%$, $D_{\mathrm{a}} = 1.36$ $\omega_{\mathrm a}/\omega_0=2$, $\mu$ between $0.4\%$ and $5\%$ depending on exciter location, leading to force drop down by $77\%$ and $97\%$, respectively)
%--> force drop, frequency shift and stronger apparent damping are not transient effects! one could counteract this, of course, using feedback control
%- it is well known [McConnell] that damping and frequency wrongly determined if applied force is not measured
%
%[fig: $|\hat F|/|\hat F_0|$ vs. $\eta$ with asymptotic limits and force drop]
%

\section{Numerical results}\label{sec:num}
% PURPOSE OF SECTION
In this section, the opportunities and limitations of the extended nonlinear-mode model are computationally assessed for the analysis of exciter-structure interaction, based on the model depicted in \fref{benchmark_model}.
A harmonic voltage input to the shaker is considered, $u(t) = \real{\hat u \ee^{\ii\int_0^t\Omega \dd \overline t}}$, with fixed magnitude $\hat u$ and linearly increasing frequency $\Omega(t)$ (\emph{frequency sweep}).
The frequency range around the lowest-frequency bending mode is studied.
\\
% PARAMETERS OF STRUCTURAL MODEL
For the structural model, the following geometrical and material parameters are specified: $\ell = 0.53~\mathrm{m}$, $h = 0.019~\mathrm{m}$, $w = 0.014~\mathrm{m}$, $E=2.1\cdot 10^{11}~~\mathrm{N}/{\mathrm{m}}^2$, $\rho = 7.8\cdot 10^3~\mathrm{kg}/\mathrm m^3$, and the cubic spring constant is set to $k_3 = 6\cdot 10^9~\mathrm{N}/\mathrm{m}^3$.
The beam is described using Euler-Bernoulli beam finite elements with $100$ nodes along the length.
Stiffness-proportional viscous damping is specified such that the considered lowest-frequency mode (linear natural frequency $\omega_0=346~\mathrm{rad}/\mathrm{s}$) has the damping ratio $D_0=0.4\%$ in the linear case.
For computational efficiency, a Craig-Bampton reduction is applied, where the bending displacement at the tip (location of nonlinear spring) is retained as master coordinate, along with $10$ normal modes (with fixed master coordinate) are considered.
The nonlinear modal analysis is carried out using Harmonic Balance, as implemented in the Matlab tool NLvib \cite{Krack.2019}.
The harmonic truncation order is set to $H=13$ if not otherwise specified.
%The nonlinear term is evaluated using the alternating frequency-time scheme with a number of samples sufficient to exclude sampling errors \cite{Woiwode.2019}.
%
\fig[b!]{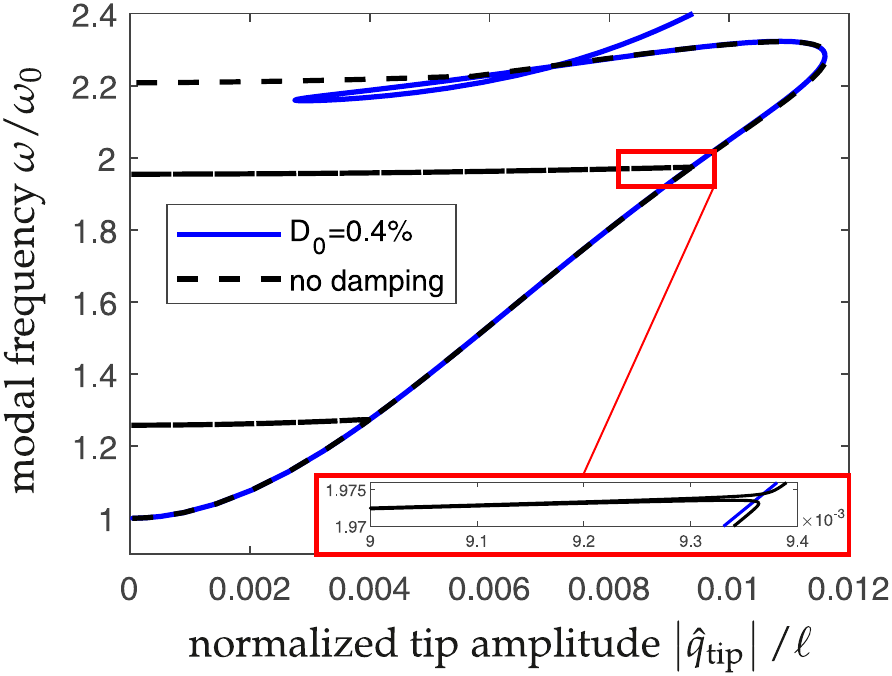}{Modal frequency of the beam with cubic spring vs. amplitude; $\left|\hat q_{\mathrm{tip}}\right|$ denotes the magnitude of the fundamental Fourier coefficient of the bending displacement at the beam's tip}{0.5}
\\
% NONLINEAR MODAL ANALYSIS RESULTS
The amplitude-dependent modal frequency of the structural model is depicted in \fref{modal_frequency_vs_amplitude}.
As amplitude measure, the magnitude of the fundamental Fourier coefficient of the bending displacement at the beam's tip is used ($\left|\hat q_{\mathrm{tip}}\right|$).
The results are shown for the underlying conservative system and the damped system ($D_0=0.4\%$).
In both cases, loops occur which can be attributed to nonlinear modal interactions.
The first, second and third loop correspond to the $1:5$ internal resonance between mode 1 and mode 2, the $1:9$ internal resonance between mode 1 and mode 3, and the $1:3$ internal resonance between mode 1 and mode 2, respectively.
%These typically occur under the condition of internal resonance of two or more nonlinear modal frequencies at the same mechanical energy level.
Along these loops, the vibration energy localizes in the higher harmonics.
Consequently, the amplitude measure, $\left|\hat q_{\mathrm{tip}}\right|$, which only considers the fundamental harmonic, approaches zero.
It is well known that these singular features are highly sensitive to damping.
Apparently, the two lower loops vanish already for $D_0=0.4\%$.
The third loop persists such that strong modal interactions can be expected to affect the dynamics of the damped system in this range.
This represents a natural limitation of the proposed nonlinear-mode model, due to the underlying assumption that the vibration energy is confined to a single nonlinear mode.
%note: conventional and EPMC coincide for this conservative case
\\
% PARAMETERS OF EXCITER MODEL
As exciter, a Br\"uel and Kjaer shaker type V4808 is considered.
Taking into account the parameters identified in \cite{Morlock2015} and the shaker's technical specifications, the parameters are set as follows: $L=140\cdot 10^{-6}~\mathrm{H}$, $R=3~\mathrm{\Omega}$, $G = 15.5~\mathrm{N}/\mathrm{A}$, $m_{\mathrm a} = 0.043\mathrm{kg}$, $k_{\mathrm a} = 2.07\cdot 10^{4}~\mathrm N/\mathrm m$, $D_{\mathrm a} = 1.36$.
Note that the armature damping is super-critical, which is here mainly due to the electro-mechanical damping (the table damping only contributes $2.36\%$ to $D_{\mathrm a} $).
One can verify that the cutoff frequency is indeed much larger than the modal frequency of interest, $\omega_{\mathrm c}/\omega_0 = 62$, as presumed in the derivation of the simplified expression of the force drop.
On the other hand, the armature-to-structure natural frequency ratio is $\omega_{\mathrm a}/\omega_0=1.998$.

\subsection{Effect of the sweep rate}
\fig[h]{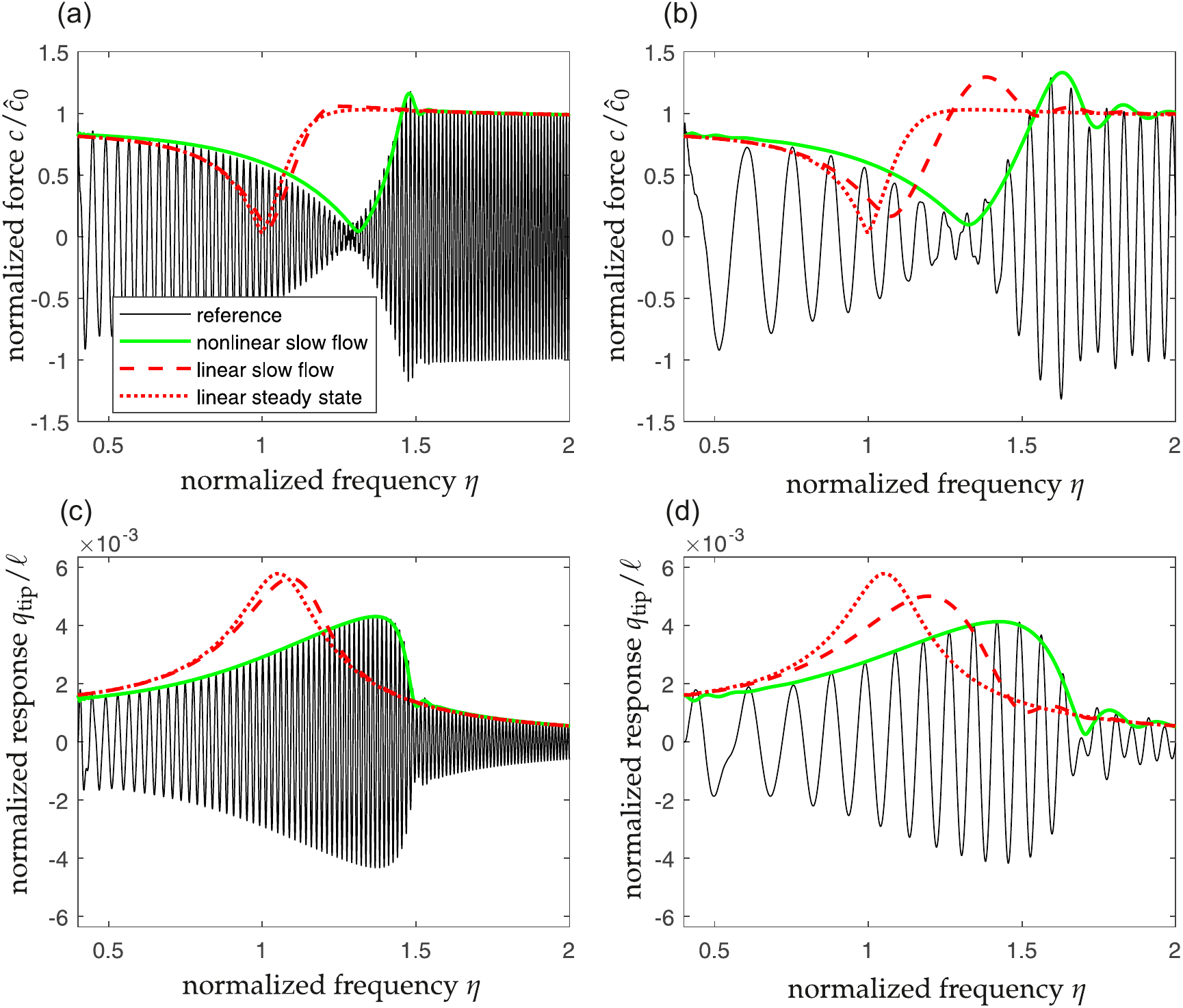}{Force (top) and displacement (bottom) during a frequency sweep with $2\%$ (left) and $10\%$ (right) frequency change per pseudo period; exciter placed at $2/3\ell$, $\hat u = 10~\mathrm V$; $\qtip$ denotes the bending displacement at the beam's tip; the legend in (a) applies to all sub-figures}{1.0}
%[fig: force and response vs. frequency/time; reference, nonlinear slow-flow, linear slow-flow, linear steady-state; for 2 different positive and one negative sweep rates;
%perhaps with zoom into synthesis with higher harmonics
%
A forward frequency sweep through the first structural resonance is studied, with a voltage level of $\hat u=10~\mathrm V$.
The exciter is placed at $2\ell/3$ (\cf \fref{benchmark_model}a).
This yields a moderate mass ratio of $\mu = 5\%$, and a substantial force drop.
In the linear steady state, the force drops by $97\%$, in accordance with \eref{forcedrop}.
\fref{varySweepRate} depicts the applied force, $\fcs$, and displacement response, $\qtip$, where $\qtip$ is the bending displacement at the beam's tip.
Recall that $\fhnull = G\hat u/R$.
The results are shown for two different sweep rates: the frequency changes by $2\%$ (\fref{varySweepRate}-left) and $10\%$ (\fref{varySweepRate}-right) per pseudo period, respectively.
Here, the reference frequency, both for the frequency change and the definition of the pseudo period is the linear natural frequency $\omega_0$.
\\
The simulation results are obtained by direct numerical integration of the assembled full-order model described by \erefs{finalassembly1}-\erefo{finalassembly2} (reference), and by numerical integration of the slow flow equation of the single modal oscillator in \eref{NM} (nonlinear slow flow).
Also depicted are the results obtained for the underlying linear system (linear slow flow, linear steady state).
This helps to distinguish nonlinear from transient effects.
The linear results are obtained by replacing the amplitude-dependent nonlinear modal properties by their respective constant linear counterparts in \eref{NM}, and then either numerical integration (linear slow flow) or setting $\dot a =0=\dot\theta$ and explicitly solving the algebraic equation (linear steady state).
\\
Away from resonance, the results obtained by all four methods approach each other.
Here, the amplitudes are relatively small such that nonlinear effects are weak.
Also, the effective dynamic stiffness (including the attachment impedance) is relatively constant such that the behavior is close to the steady limit state.
%\\
The results in \fref{varySweepRate} indicate a very high accuracy of the single-nonlinear-mode model, both with respect to the displacement response and the applied force.
Both the force drop and the subsequent force overshoot are well-captured in the reduced model.
For a faster sweep, the deviation between steady state and transient response is more pronounced.
Interestingly, the displacement amplitude decreases suddenly in a range where the applied force is relatively constant.
This is the well-known jump phenomenon, typical for frequency sweeps of lightly-damped systems with nonlinear stiffness properties (here: Duffing-type hardening).
Even for the rather fast sweep in \fref{varySweepRate}-right, the force and displacement amplitude modulation is approximated with reasonable accuracy.
The remaining deviations can be attributed to two aspects:
First, considerable higher harmonics occur, in particular in the applied force.
Here, the accuracy is limited as \eref{NMpre} is derived by projecting onto only the fundamental harmonic component of the nonlinear mode.
This is seen in an even more pronounced way in the next subsection.
Second, the time scales of amplitude-phase-modulation and vibration are relatively close in the quicker sweep.
Here, the assumption of separate time scales, underlying the averaging procedure to derive \eref{NMpre}, is violated.
%nominal parameters in benchmark: $D_0=0.4\%$, $D_{\mathrm{a}} = 1.36$ $\omega_{\mathrm a}/\omega_0=2$, $\mu$ between $0.4\%$ and $5\%$ depending on exciter location, leading to force drop down by $77\%$ and $97\%$, respectively)

\subsection{Effect of excitation level and occurrence of higher harmonics}
\fig[h]{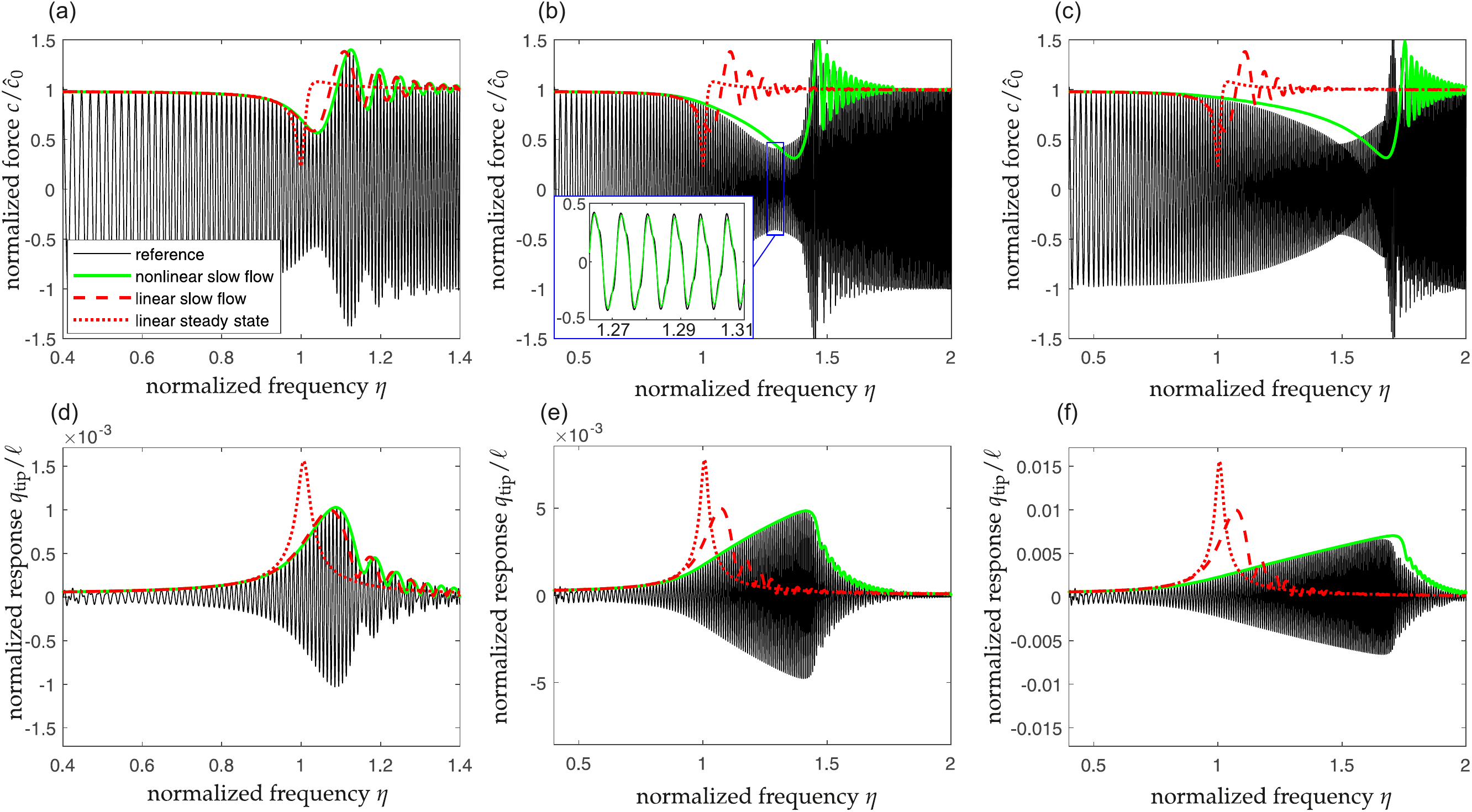}{Force (top) and displacement (bottom) during a frequency sweep with voltage level $1~\mathrm V$ (left), $5~\mathrm V$ (center), and $10~\mathrm V$ (right);
exciter placed at $1/3\ell$; $1\%$ frequency change per pseudo period; the legend in (a) applies to all sub-figures}{1.0}
Now, the exciter is placed at $\ell/3$ (\cf \fref{benchmark_model}a).
As the modal deflection is smaller here, as compared with the location $2\ell/3$, the modal mass is larger.
This yields a small mass ratio of $\mu = 0.4\%$, and a less pronounced force drop.
In the linear steady state, the force drops by $77\%$, in accordance with \eref{forcedrop}.
The sweep rate is now set to have a $1\%$ frequency change per pseudo period.
Three different voltage levels are considered: $1~\mathrm V$, $5~\mathrm V$, and $10~\mathrm V$.
The results are depicted in \fref{varyExcitationLevel} left, center, and right, respectively.
\\
Even in the linear case, the force drop is less pronounced during the transient sweep as compared with the steady limit state.
Also, the force amplitude shows a strong modulation after crossing the structural resonance.
In the nonlinear case, the results reflect the expected hardening characteristic and skewness of the displacement envelope.
With increased voltage level, the nonlinearity becomes more important, and thus higher harmonics are more pronounced.
This can be seen, for instance, in the zoom within \fref{varyExcitationLevel}b.
As discussed before, the nonlinear-mode approximation is less accurate in this case.
Still, the displacement response amplitude and even the higher harmonic content of the force, determined in the post-processing using \erefs{ansatz} and \erefo{fc1}, are captured with reasonable accuracy.
% Mainly the time-evolution of the modal amplitude and phase appear to deviate as the excitation level increases.

\subsection{Sensitivity of nonlinear-mode model to frequency errors in the case of imposed dynamic forcing}
\fig[h]{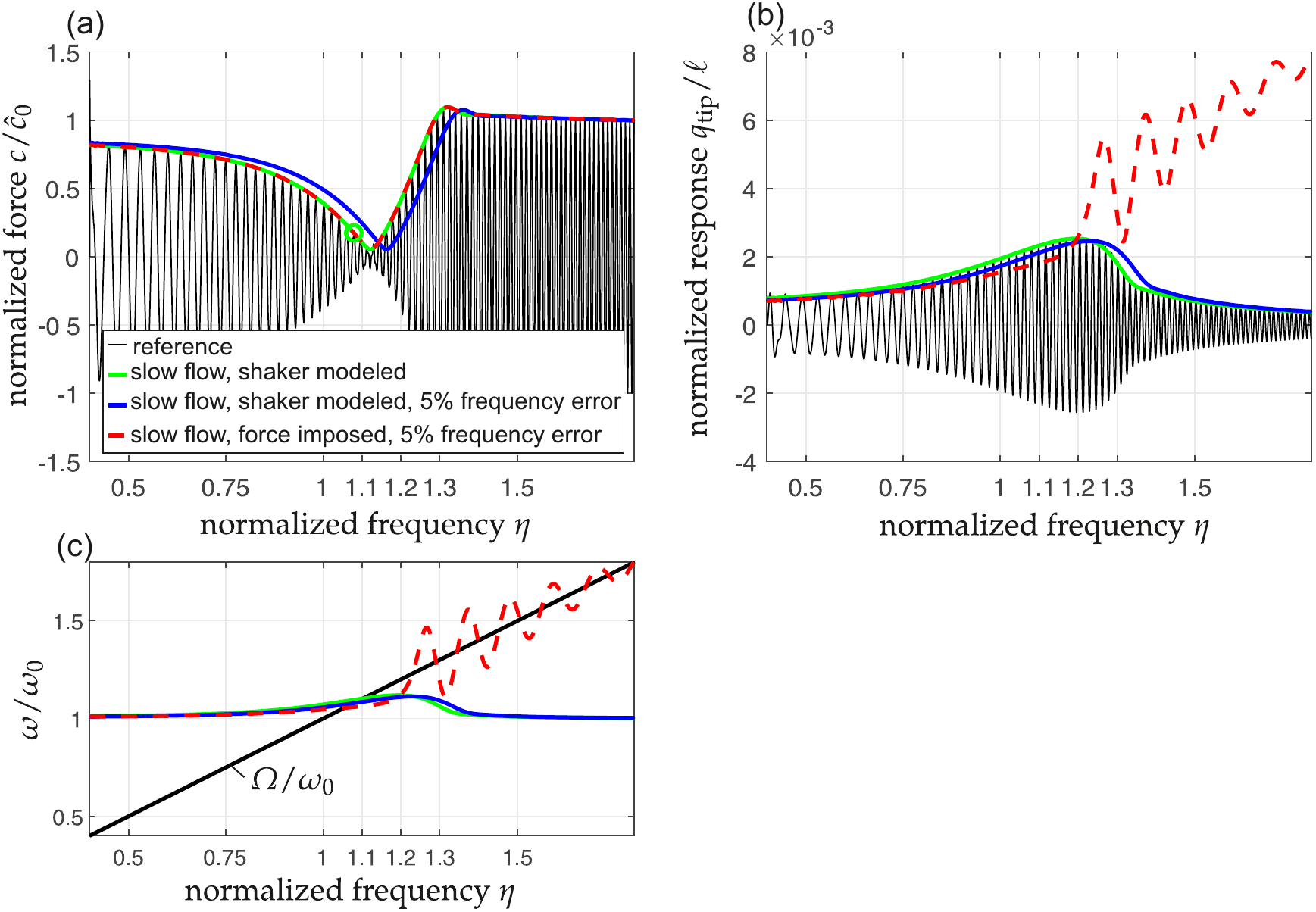}{Effect of natural frequency error on force (a) and displacement (b) during a frequency sweep; (c) illustrates the natural frequency and the excitation frequency; exciter placed at $2/3\ell$, $\hat u = 5~\mathrm V$, $2\%$ frequency change per pseudo period; the legend in (a) applies to all sub-figures}{1.0}
%[fig: force and response vs. frequency/time; ... both linear AND nonlinear]
%
It is common practice to assess the quality of the nonlinear-mode model by comparison to directly measured stepped or swept frequency responses.
It seems natural to request that the nonlinear-mode model accurately predicts the response to frequency sweeps through resonance.
However, relatively poor prediction accuracy and even unbounded amplitude growth was found in \cite{Scheel.2020}.
The theoretical developments proposed in this work permit to explain this unexpected behavior, as discussed in this subsection.
It should be emphasized that such natural frequency errors are inevitable, regardless of whether the nonlinear modal analysis is carried out experimentally or computationally (based on a suitable model).
For lack of an exciter model, one might wish to measure the force in the sweep test, and impose this force in the nonlinear-mode model.
With this, one obviously neglects the exciter-structure interaction and, thus, the causal relation leading to the force drop, as was done in \cite{Scheel.2020}.
As will be shown, \emph{if the exciter-structure interaction is ignored}, the nonlinear-mode model becomes highly sensitive to natural frequency errors.
\\
\fref{effectModalStiffnessErrorImposedForce}a-b depicts the force and displacement response determined by the full-order model (reference) and three different variants of the nonlinear-mode model:
(1) with a shaker model, (2) with a shaker model and a $+5\%$ natural frequency error, (3) with imposed force and a $+5\%$ natural frequency error.
The error is introduced as an amplitude-independent offset.
The imposed force for configuration (3) was adopted from the nonlinear-mode model (1) here.
For completeness, the imposed force was also determined from the reference.
Thanks to the excellent agreement of the reference force envelope with the results of (1), this lead to practically indistinguishable results, which are not shown here for clarity of the figure.
Moreover, it should be remarked that the configuration with imposed force but no frequency error yields the same results as configuration (1), which are also not shown here for clarity of the figure.
In \fref{effectModalStiffnessErrorImposedForce}c, the normalized natural frequency, $\omega/\omega_0$, and the normalized excitation frequency, $\Omega/\omega_0$, are illustrated as well.
Since $\eta=\Omega/\omega_0$, the normalized excitation frequency is a straight line in this sub-figure and just depicts the identity.
In contrast, the natural frequencies depend on the amplitude, which, in turn, varies with time and thus $\eta$.
In \fref{effectModalStiffnessErrorImposedForce}c is useful for locating the points where the excitation frequency coincides with the natural frequency, and thus the resonance condition is met.
\\
The nonlinear-mode model with natural frequency offset and imposed force sees the same applied force, but responds quite differently than the other models.
The reason is that at the point where the excitation frequency coincides with the (erroneous) natural frequency, the applied force level is not near the minimum of the envelope, but at a level on the steep slope beyond this minimum (\cf \fref{effectModalStiffnessErrorImposedForce} c and a).
Consequently, the nonlinear-mode model responds with much larger amplitudes, which quickly exceed the model's validity limit.
In contrast, when the shaker is modeled, with the approach proposed in this work, a small natural frequency error yields only small response errors.
It should be emphasized, again, that the described phenomenon is neither a transient nor a nonlinear effect.
Finally, it is useful to note that changing the sign of the offset, from $+5\%$ to $-5\%$, does not necessarily lead to a qualitative change of the effect.
In that case, the offset resonance is reached earlier, at a force level on the steep slope before the minimum of the force envelope.
\\
% COMPUTATION EFFORT
Using the explicit variable time step Runge-Kutta Dormand-Prince method (Matlab ode45, default tolerances), the computation times to obtain \fref{effectModalStiffnessErrorImposedForce} on the author's personal computer are as follows: The nonlinear modal analysis (truncation order reduced to $H=3$ as found sufficient) requires $0.1~\mathrm{s}$ for the required amplitude range, the numerical integration of the reference model requires $13\mathrm{s}$, and the numerical integration of each nonlinear-mode model requires $1.3\mathrm{s}$.
It should be remarked that the computation effort depends on many aspects.
In particular, the initial model order has a crucial influence on the computation effort of the reference, whereas the dimension of the single-nonlinear-mode model is independent of this.
% It should be noted that in this particular example, the finite stiffness of the stinger was modeled with $k_{\mathrm s} = 2.6\cdot 10^7~\mathrm{N/m}$.

\section{Conclusions}\label{sec:conc}
% THEORY DEVELOPMENT
In the present work, the single-nonlinear-mode theory was extended to account for linear attachments.
The linear attachments must be described by linear ordinary differential or differential-algebraic equation systems with constant coefficient matrices.
Assuming that the modal deflection shape is not significantly distorted by the attachments, their effect can be reduced to a complex-valued modal impedance and an imposed modal forcing.
These reduced modal terms are derived in the frequency domain using the framework of dynamic substructuring.
In the transient case, the amplitude and phase modulation can still be predicted by the single nonlinear modal oscillator model alone, if the attachment is assumed to be in quasi-steady state.
Otherwise, the modulation equations would have to be extended by the attachment states.
\\
% ASSESSMENT FOR EXCITER-STRUCTURE INTERACTION
The proposed approach was then applied to exciter-structure interaction, considering a cantilevered beam with cubic spring and an established electro-mechanical shaker model.
The near-resonant force drop encountered during frequency response testing was briefly revisited.
The results indicate that the nonlinear-mode model is highly accurate in a wide range.
Besides the aforementioned limitation that the attachments must not significantly distort the modal deflection shape, only the already well-known limitations of the single-nonlinear-mode theory reappeared.
In particular, strongly nonlinear modal interactions are assumed to be absent.
Inaccuracy must also be expected in the presence of strong higher harmonics.
This is the price to pay for the simplified projection involving the fundamental harmonic only, which is the key to expressing the nonlinear terms in closed form (hyper-reduction).
In the transient case, the time scales of amplitude and phase modulation on the one hand and the vibration on the other hand must be well-separated.
Finally, it is shown that if the exciter is not modeled but replaced by an imposed force, small modal frequency errors may cause large response errors or even divergence.
If the shaker is modeled as attachment, this high sensitivity vanishes.
This underlines the importance to model the exciter, in particular to properly predict uncontrolled sweeps through resonance.
\\
% IMPLICATIONS / PERSPECTIVE
The extension of single-nonlinear-mode theory by linear attachments considerably extends the range of utility of nonlinear modes.
Potential application fields are diverse, and may include structural modification, vibration energy harvesting, active/smart structures, or controller design.

\end{document}